\documentclass[preprint,onecolumn,showpacs,amsmath,amssymb,prd]{revtex4}
\usepackage{epsf}
\usepackage{dcolumn}
\usepackage{bm}
\everymath{\displaystyle}
\begin{document}

\title{New symmetry properties of pointlike scalar and Dirac particles}
\author{Alexander J. Silenko}
\affiliation{Research Institute for Nuclear Problems, Belarusian State University, Minsk 220030, Belarus\\
Bogoliubov Laboratory of Theoretical Physics, Joint Institute for Nuclear Research, Dubna 141980, Russia}

\date{\today}

\begin{abstract}
New symmetry properties are found for pointlike scalar and Dirac particles (Higgs boson and all leptons) in Riemannian and Riemann-Cartan spacetimes in the presence of electromagnetic interactions. A Hermitian form of the Klein-Gordon equation for a pointlike scalar particle in an arbitrary \emph{n}-dimensional Riemannian (or Riemann-Cartan) spacetime is obtained. New conformal symmetries of initial and Hermitian forms of this equation are ascertained. In the above spacetime, general Hamiltonians in the generalized Feshbach-Villars and Foldy-Wouthuysen representations are derived. The conformal-like transformation conserving these Hamiltonians is found. Corresponding conformal symmetries of a Dirac particle are determined. It is proven that all conformal symmetries remain unchanged by an inclusion of electromagnetic interactions.
\end{abstract}

\pacs {11.30.-j, 03.65.Pm, 04.62.+v, 11.10.Ef}
\maketitle

\section{Introduction}

A determination of symmetry properties of elementary particles is one of the most important problems of contemporary particle physics. Symmetries of basic relativistic wave equations describing pointlike particles with spin 0 (Higgs boson) and 1/2 (all leptons) retain an important place among these properties. Intensive studies of such symmetries began fifty years ago with the seminal work by Penrose \cite{Penrose}. He discovered the conformal invariance of the
covariant Klein-Gordon (KG) equation \cite{KG} for a
massless scalar particle in Riemannian spacetimes and supplemented this equation with
a term describing a nonminimal coupling to the scalar
curvature. Chernikov and Tagirov
\cite{CheTagi} have studied the case of a nonzero mass and
$n$-dimensional Riemannian spacetimes. The inclusion of the additional
Penrose-Chernikov-Tagirov (PCT) term has been argued for both massive and massless particles
\cite{CheTagi}. Accioly and Blas
\cite{AccBlas} have performed the exact
Foldy-Wouthuysen (FW) transformation for a massive spin-0 particle in static
spacetimes and have found new telling arguments in favor of the predicted
coupling to the scalar curvature. A derivation of the relativistic
FW Hamiltonian is very important for a comparison of gravitational (and inertial) effects in classical and quantum gravity
because the FW representation restores Schr\"{o}dinger-like forms
of Hamiltonians and equations of motion. These forms are
convenient for finding a semiclassical approximation and
a classical limit of relativistic quantum mechanics (see Refs. \cite{workFWt,JMP,PRD,PRA,PRAnonstat} and references therein).

However, the transformation method used in Ref. \cite{AccBlas} is
not applicable to either massless particles or
nonstatic spacetimes. To find a specific
manifestation of the conformal
invariance in the FW representation which takes place just for massless particles, the generalized Feshbach-Villars (GFV)
transformation \cite{TMP2008} applicable for such particles have been performed \cite{Honnefscalar}.
The subsequent relativistic FW transformations has made it possible to derive the FW Hamiltonians for the
both massive and massless scalar particles in general noninertial
frames and stationary gravitational fields. The new manifestation of the conformal
invariance for massless particles consisting in the conservation of the FW Hamiltonian and
the FW wave function has been discovered. New exact FW Hamiltonians have been obtained for both massive and massless scalar particles
in general static spacetimes and in frames rotating in the Kerr field approximated by a spatially isotropic metric.
The high-precision expression for the FW Hamiltonian has been derived in the general case. It has also been shown that
conformal transformations change only such terms in the FW Hamiltonians that are proportional to the particle mass $m$.

In the present work, we consider the much more general problem of scalar and Dirac particles in arbitrary
gravitational (noninertial) and electromagnetic fields and find (on a quantum-mechanical level)
new symmetry properties relative to conformal transformations not only in the FW representation but also in initial
representations. These properties are attributed to all known pointlike scalar and Dirac particles (Higgs boson and leptons) and also to the hypothetic pseudoscalar axion.

We denote world and spatial indices by Greek and Latin letters
$\alpha,\mu,\nu,\ldots=0,1,2,3,~i,j,k,\ldots=1,2,3$, respectively.
Tetrad indices are denoted by Latin letters from the beginning of the
alphabet, $a,b,c,\ldots = 0,1,2,3$. 
Temporal and spatial tetrad indices are
distinguished by hats.
The signature is $(+---)$, and the Ricci scalar
curvature is defined by
$R=g^{\mu\nu}R_{\mu\nu}=g^{\mu\nu}R^\alpha_{~\mu\alpha\nu}$, where
$R^\alpha_{~\mu\beta\nu}=\partial_\beta\Gamma^\alpha_{~\mu\nu}-\ldots$
is the Riemann curvature tensor. The denotation $f_{,\mu}$ means $\partial f/(\partial x^\mu)$. We use the system of units
$\hbar=1,~c=1$. 

\section{Hermitian form of the Klein-Gordon equation and conformal
symmetry for a pointlike scalar particle}\label{HfKGe}

The covariant KG 
equation with the additional PCT term \cite{Penrose,CheTagi} describing a scalar particle in
an $n$-dimensional Riemannian spacetime is given by
\begin{equation}
(\square+m^2-\lambda R)\psi=0,~~~
\square\equiv\frac{1}{\sqrt{-g}}\partial_\mu\sqrt{-g}g^{\mu\nu}\partial_\nu.
\label{eqKG} \end{equation} Minimal (zero) coupling corresponds
to $\lambda=0$, while the PCT coupling is defined by $\lambda=(n-2)/[4(n-1)]$
\cite{CheTagi}. The sign of the Penrose-Chernikov-Tagirov term depends on the definition of $R$.
For noninertial (accelerated and rotating) frames, the spacetime is flat and $R=0$.

For a \emph{massless} particle, the conformal transformation
\begin{equation}
\widetilde{g}_{\mu\nu}=O^{-2}g_{\mu\nu}\label{conftrf}
\end{equation} conserves
the form of Eq. (\ref{eqKG}) but changes the wave
function and the operators acting on it \cite{Penrose,CheTagi}:
\begin{equation}
\square-\lambda R=O^{-\frac{n+2}{2}}\bigl(\widetilde{\square}-\lambda \widetilde{R}\bigr)O^{\frac{n-2}{2}},~~~
\widetilde{\psi}=O^{\frac{n-2}{2}}\psi. \label{eqcin}
\end{equation}

To specify symmetry properties of the initial KG 
equation (\ref{eqKG}), it is instructive to present it in the Hermitian form. Amazingly, this can be achieved with the simple nonunitary transformation
\begin{equation}
\psi=f^{-1}\Phi,~~~ f=\sqrt{g^{00}\sqrt{-g}},~~~ g=\det{g_{\mu\nu}}. \label{nonunKG}
\end{equation} Since $\widetilde{g}=O^{-2n}g$, $\Phi$ is invariant relative to the conformal transformation (\ref{conftrf}). This invariance takes place only for a massless particle. After the transformation (\ref{nonunKG}), we multiply the obtained equation by the factor $f/g^{00}$ and come to the Hermitian form of the KG equation:
\begin{equation}
\left(\frac{1}{f}\partial_\mu\sqrt{-g}g^{\mu\nu}\partial_\nu\frac{1}{f}+\frac{m^2}{g^{00}}-\frac{\lambda R}{g^{00}}\right)\Phi=0.
\label{eqKGH} \end{equation}

The use of Eqs. (\ref{conftrf})--(\ref{nonunKG}) shows that Eq. (\ref{eqKGH}) is conformally invariant for a massless particle. However, it is not conformally invariant for a massive one. To determine its conformal symmetry in the latter case, it is sufficient to find a physical quantity that when substituted for $m$ restores the conformal invariance of Eq. (\ref{eqKGH}). For this purpose, we can use the quantity $m'$ which is equal to $m$ in the initial spacetime and takes the form \begin{equation}\widetilde{m'}=Om' \label{eqconfm} \end{equation} after the conformal transformation (\ref{conftrf}). The equation obtained from Eq. (\ref{eqKGH}) with the substitution of $m'$ for $m$,
 \begin{equation}
\left(\frac{1}{f}\partial_\mu\sqrt{-g}g^{\mu\nu}\partial_\nu\frac{1}{f}+\frac{{m'}^2}{g^{00}}-\frac{\lambda R}{g^{00}}\right)\Phi=0,
\label{eqKGHcf} \end{equation}
is conformally invariant. While this equation does not describe a real particle and is not equivalent to Eq. (\ref{eqKGH}), finding the appropriate substitution (\ref{eqconfm}) determines the conformal symmetry of the suitable equation (\ref{eqKGH}). The determination of a new symmetry property for massive particles is rather important because the only discovered pointlike scalar particle, the Higgs boson, is massive.

Thus, we can conclude that Eq. (\ref{eqKGH}) is not changed by the conformal-like transformation
\begin{equation}
\widetilde{g}_{\mu\nu}=O^{-2}g_{\mu\nu}, ~~~ m\rightarrow m', ~~~ \widetilde{m'}=Om'. \label{conflike}
\end{equation}  In particular, this transformation does not change the wave function $\Phi$. In the general case, we can substitute any quantity satisfying Eq. (\ref{eqconfm}) for $m$ into Eq. (\ref{eqKGH}).

We can now state the conformal symmetry of the initial KG equation (\ref{eqKG}). The substitution of $m'$ for $m$ makes the \emph{changed} equation conformally invariant with the following properties:
\begin{equation} \begin{array}{c}
\square+{m'}^2-\lambda R=O^{-\frac{n+2}{2}}\bigl(\widetilde{\square}+{\widetilde{m'}}^2-\lambda \widetilde{R}\bigr)O^{\frac{n-2}{2}},\\
\widetilde{\psi}=O^{\frac{n-2}{2}}\psi. \end{array} \label{eqcintr}
\end{equation} These properties establish the conformal symmetry of the covariant KG
equation (\ref{eqKG}) and the specific form of its invariance relative to the conformal-like transformation (\ref{conflike}).

The method of the FW transformation used in Ref. \cite{Honnefscalar} is applicable to nonstationary spacetimes. However, only the stationary case has been considered in this work. To make a more general investigation of 
symmetry properties in the FW representation, we need to present Eq. (\ref{eqKGH}) in another (equivalent) form.

Let us introduce the following denotations:
\begin{equation} \begin{array}{c}
\Gamma^i=\sqrt{-g}g^{0i}, ~~~ G^{ij}=g^{ij}-\frac{g^{0i}g^{0j}}{g^{00}}. \end{array} \label{denoton}
\end{equation}
Lengthy but straightforward calculations bring Eq. (\ref{eqKGH}) to the form
\begin{equation}
\left[(\partial_0+\Upsilon)^2+\partial_i\frac{G^{ij}}{g^{00}}\partial_j+\frac{m^2}{g^{00}}+\Lambda\right]\Phi=0,
\label{eqKGHnew} \end{equation} where
\begin{equation} \begin{array}{c}
\Upsilon=\frac{1}{2f}\left\{\partial_i,\Gamma^i\right\}\frac{1}{f}=\frac{1}{2}\left\{\partial_i,\frac{g^{0i}}{g^{00}}\right\}, \\ \Lambda=-\frac{f_{,0\,,0}}{f}-\left(\frac{g^{0i}}{g^{00}}\right)_{,i}\frac{f_{,0}}{f}-2\frac{g^{0i}}{g^{00}}\frac{f_{,0\,,i}}{f}-
\left(\frac{g^{0i}}{g^{00}}\right)_{,0}\frac{f_{,i}}{f}
\\-\frac{1}{2}\left(\frac{g^{0i}}{g^{00}}\right)_{,0\,,i}-\frac{1}{2f^2}\left(\frac{g^{0i}}{g^{00}}\right)_{,i}\Gamma^j_{,j}-
\frac{g^{0i}}{2f^2g^{00}}\Gamma^j_{,j\,,i}\\
+\frac{1}{4f^2}\left(\Gamma^i_{,i}\right)^2-\left(\frac{G^{ij}}{g^{00}}\right)_{,i}\frac{f_{,j}}{f}-\frac{G^{ij}}{g^{00}}\frac{f_{,i\,,j}}{f}
-\frac{\lambda R}{g^{00}}. \end{array} \label{denotatin}
\end{equation}

This form of the KG equation is also Hermitian and the wave function is not changed as compared with Eq. (\ref{eqKGHcf}). The replacement of $m$ with $m'$ makes Eq. (\ref{eqKGHnew}) to be conformally invariant. Therefore, Eq. (\ref{eqKGHnew}) is invariant relative to the conformal-like transformation (\ref{conflike}).

\section{Conformal symmetries of Hamiltonians}\label{confsym}

To fulfill the successive GFV and
FW transformations, we use the method developed in Ref. \cite{TMP2008} and applied to the covariant KG equation in Ref. \cite{Honnefscalar}. The original Feshbach-Villars method does not work for massless particles while its generalization \cite{TMP2008} makes it possible to extend the method to such particles.

We introduce two new functions, $\phi$ and $\chi$, as follows \cite{TMP2008,Honnefscalar}:
\begin{equation} \begin{array}{c} \Phi=\phi+\chi, ~~~
i\left(\partial_0+\Upsilon\right)\Phi= N(\phi-\chi),
\end{array} \label{eq3i} \end{equation} where $N$ is an arbitrary nonzero real
parameter. For the Feshbach-Villars transformation, it is definite
and equal to the particle mass $m$. These functions form the two-component wave function in the GFV representation, $\Psi=\left(\begin{array}{c} \phi \\ \chi \end{array}\right)$. Equations (\ref{eqKGHnew}) and (\ref{eq3i}) result in (cf. Ref. \cite{Honnefscalar})
\begin{equation} \begin{array}{c}  i\frac{\partial\Psi}{\partial t}={\cal H}\Psi,
~~~ {\cal
H}=\rho_3
\frac{N^2+T}{2N}\!+\!i\rho_2 \frac{-N^2+T}{2N}-i\Upsilon,\\
 T=\partial_i\frac{G^{ij}}{g^{00}}\partial_j+\frac{m^2}{g^{00}}+\Lambda,
\end{array} \label{eq5i} \end{equation}
where ${\cal H}$ is the GFV Hamiltonian and
$\rho_i~(i=1,2,3)$ are the Pauli matrices. Equation (\ref{eq5i}) is exact.

For a massless particle, this Hamiltonian is not changed by the conformal transformation (\ref{conftrf}). In the general case, it is invariant relative to the conformal-like transformation (\ref{conflike}).


The general methods developed in Refs. \cite{JMP,PRA,TMP2008} allow us to perform the FW transformation of the Hamiltonian (\ref{eq5i}) for a relativistic
particle in external fields. These methods are iterative. The initial Hamiltonian can be presented in the general form
\begin{equation} \begin{array}{c} {\cal H}=\rho_3
{\cal M}+{\cal E}+{\cal O}, ~~~\rho_3 {\cal M}={\cal M}\rho_3,\\
\rho_3 {\cal E}={\cal
E}\rho_3, ~~~\rho_3 {\cal O}=-{\cal O}\rho_3,
\end{array}
\label{eqH} \end{equation}
where ${\cal E}$ and ${\cal O}$ denote the sums of even (diagonal) and odd (off-diagonal) operators, respectively. In the considered case, $[{\cal M},{\cal O}]=0$,
\begin{equation} {\cal M}=\frac{N^2+T}{2N}, ~~~ {\cal E}=-i\Upsilon, ~~~ {\cal
O}=i\rho_2 \frac{-N^2+T}{2N},
\label{eqdH} \end{equation} and the transformation operator found in Ref. \cite{PRA} reduces to the form \cite{TMP2008,Honnefscalar}
\begin{equation}
U=\frac{\epsilon+N+\rho_1(\epsilon-N)}{2\sqrt{\epsilon N}},
~~~\epsilon=\sqrt{{\cal M}^2+{\cal O}^2}=\sqrt{T}. \label{eq8i}
\end{equation} This transformation operator is $\rho_3$-pseudounitary ($U^\dag=\rho_3U^{-1}\rho_3$).

It is important that the Hamiltonian obtained as a result of the transformation with the operator
(\ref{eq8i}) does not depend on $N$ \cite{TMP2008}:
\begin{equation} \begin{array}{c}
{\cal H}'=\rho_3\epsilon+{\cal E}'+{\cal
O}',~~~\rho_3{\cal E}'={\cal E}'\rho_3,~~~ \rho_3{\cal O}'=-{\cal O}'\rho_3,\\ {\cal
E}'=-i\Upsilon+\frac{1}{2\sqrt{\epsilon}}\left[{\sqrt{\epsilon},
[\sqrt{\epsilon}},{\cal F}]\right]\frac{1}{\sqrt{\epsilon}}, \\
{\cal O}'=\!\rho_1\frac{1}{2\sqrt{\epsilon}}[\epsilon,{\cal
F}]\frac{1}{\sqrt{\epsilon}},~~~ {\cal F}=-i\partial_0-i\Upsilon.
\end{array}\label{eq12i}\end{equation}
This shows a self-consistency of the used transformation method. The \emph{exact} intermediate Hamiltonian (\ref{eq12i}) describes massive and massless particles and is not changed by the conformal-like transformation (\ref{conflike}).

The next transformation \cite{TMP2008} eliminates residual odd terms and leads to the final form of the
\emph{approximate} relativistic FW Hamiltonian:
\begin{equation} \begin{array}{c}
{\cal H}_{FW}=\rho_3\epsilon-i\Upsilon-\frac{1}{2\sqrt{\epsilon}}\left[\sqrt{\epsilon},
\left[\sqrt{\epsilon},(i\partial_0+i\Upsilon)\right]\right]\frac{1}{\sqrt{\epsilon}}.
\end{array} \label{eqf} \end{equation} This final Hamiltonian is also invariant relative to the conformal-like transformation (\ref{conflike}). As a rule, the
relativistic FW Hamiltonian is expanded in powers of the Planck constant and is useful when the de Broglie wavelength is much smaller than the characteristic distance \cite{PRA}. In such a Hamiltonian, terms proportional to the zeroth and first powers of the Planck constant are determined exactly while higher-order terms are not specified (see Ref. \cite{FWproof}). As a result, the last term in Eq. (\ref{eqf}) can be omitted if it is proportional to the second or higher orders of $\hbar$.

\section{Inclusion of electromagnetic interactions}\label{KGelm}

Fortunately, an inclusion of electromagnetic interactions does not lead to any significant complication of the above derivations. The initial covariant KG equation takes the form
\begin{equation}
\left[g^{\mu\nu}(\nabla_\mu+ieA_\mu)(\nabla_\nu+ieA_\nu)+m^2-\lambda R\right]\psi=0,
\label{eqKGe} \end{equation}
where $\nabla_\mu$ is the covariant derivative and $A_\mu$ is the electromagnetic field potential. This equation is equivalent to the following one:
\begin{equation}
\left(\frac{1}{\sqrt{-g}}D_\mu\sqrt{-g}g^{\mu\nu}D_\nu
+m^2-\lambda R\right)\psi=0,
\label{eqKGeqv} \end{equation} where $D_\mu=\partial_\mu+ieA_\mu$.
The nonunitary transformation (\ref{nonunKG}) brings it to the Hermitian form corresponding to Eq. (\ref{eqKGH}):
\begin{equation}
\left(\frac{1}{f}D_\mu\sqrt{-g}g^{\mu\nu}D_\nu\frac{1}{f}+\frac{m^2}{g^{00}}-\frac{\lambda R}{g^{00}}\right)\Phi=0.
\label{eqKGHem} \end{equation}

It is convenient to present this equation in the equivalent form [cf. Eq. (\ref{eqKGHnew})]
\begin{equation} \begin{array}{c}
\left[(D_0+\Upsilon')^2+D_i\frac{G^{ij}}{g^{00}}D_j+\frac{m^2}{g^{00}}+\Lambda\right]\Phi=0,\\
\Upsilon'=\frac{1}{2f}\left\{\partial_i,\Gamma^i\right\}\frac{1}{f}=\frac{1}{2}\left\{D_i,\frac{g^{0i}}{g^{00}}\right\},\\
 T'=D_i\frac{G^{ij}}{g^{00}}D_j+\frac{m^2}{g^{00}}+\Lambda,\end{array}
\label{eqKGHnge} \end{equation} where
$G^{ij}$ and $\Lambda$ are defined by Eqs. (\ref{denoton}) and (\ref{denotatin}), respectively.

A repeat of the transformation given above allows us to derive the Hamiltonian in the GFV representation:
\begin{equation} \begin{array}{c}  
{\cal H}=\rho_3\frac{N^2+T'}{2N}+\rho_2 \frac{-N^2+T'}{2N}-i\Upsilon'+eA_0.
\end{array} \label{eq5e} \end{equation}

The FW transformation can be fulfilled with the operator
(\ref{eq8i}) where $\epsilon=\sqrt{T'}$. The transformed operator
${\cal H}'$ is independent of $N$. The final approximate FW
Hamiltonian is given by ($\epsilon=\sqrt{T'}$)
\begin{equation} \begin{array}{c}
{\cal H}_{FW}=\rho_3\epsilon-i\Upsilon'+eA_0\\-\frac{1}{2\sqrt{\epsilon}}\left[\sqrt{\epsilon},
\left[\sqrt{\epsilon},(i\partial_0+i\Upsilon'-eA_0)\right]\right]\frac{1}{\sqrt{\epsilon}}.
\end{array} \label{eqfem} \end{equation}
The last term in Eq. (\ref{eqfem}) can be omitted if it is
proportional to the second or higher orders of $\hbar$ (see
previous section).

All Hamiltonians obtained with the inclusion of electromagnetic interactions (${\cal H},~{\cal H}'$, and ${\cal H}_{FW}$) are invariant relative to the conformal-like transformation (\ref{conflike}). The Hamiltonians are conformally invariant for a massless particle. Thus, this inclusion does not change the conformal symmetries of the Hamiltonians.

In Secs. \ref{HfKGe}--\ref{KGelm}, we considered a scalar particle in Riemannian spacetimes. However, all results obtained remain applicable to Riemann-Cartan spacetimes. The spacetime torsion does not affect the Hamiltonian of a massless particle and the corresponding equation of motion. In particular, the torsion couples only to the particle spin and is not attached to the orbital angular momentum of a test particle \cite{HehlObukhovPuetzfeld}.

\section{Conformal symmetry properties of Dirac particles}

It is easy to determine the conformal symmetry properties of a pointlike Dirac particle. It has been established in Ref. \cite{Honnefscalar} that the Dirac and FW Hamiltonians for a \emph{massless} particle and the corresponding wave functions are invariant relative to the conformal transformation (\ref{conftrf}). The initial covariant Dirac equation is also conformally invariant relative to this transformation. After the conformal transformation (\ref{conftrf}), the wave function of the  Dirac equation for a \emph{massless} particle acquires the additional factor $O^{3/2}$ \cite{Honnefscalar}.

These results can be extended to massive particles. A pointlike particle in Riemannian spacetimes is described by the covariant Dirac equation
(see Refs. \cite{HehlTwoLectures,OSTRONG} and references therein)
\begin{equation}
\begin{array}{c}
\left(i\hbar \gamma^a D_a - mc\right)\psi = 0,~~~D_a = e_a^\mu D_\mu,\\
D_\mu = \partial _\mu + ieA_\mu + {\frac i4}\sigma^{ab}\Gamma_{\mu\,ab},
\end{array}\label{Dirac0}
\end{equation}
where $D_\mu$ is the covariant derivative, $\sigma^{ab} = i\left(\gamma^a \gamma^b - \gamma^b\gamma^a
\right)/2$, and the Dirac matrices $\gamma^a$ are defined in local Lorentz (tetrad)
frames. The anholonomic components of the connection are \cite{HehlTwoLectures,OSTRONG}
\begin{equation}
\Gamma_{\mu\,ab} =-\Gamma_{\mu\,ba} = e_\mu^c \Gamma_{cab},~~~ \Gamma_{cab}=\frac12\left(-C_{cab}+C_{abc}-C_{bca}\right), ~~~
C_{abc}=e_a^\mu e_b^\nu(e_{c\nu,\mu}-e_{c\mu,\nu}),\label{eqnnn}
\end{equation} where $e_\mu^a$ is the tetrad and $e_a^\mu$ is the inverse
tetrad. It is convenient to parametrize the spacetime metric as follows \cite{OSTRONG}:
\begin{equation}\label{LT}
ds^2 = V^2(dx^0)^2 - \delta_{\widehat{i}\widehat{j}}W^{\widehat
i}{}_k W^{\widehat j}{}_l\,(dx^k - K^kdx^0)\,(dx^l - K^ldx^0).
\end{equation}
The functions $V$ and $K^i$, as well as the components of the
$3\times 3$ matrix $W^{\widehat i}{}_j$ may depend arbitrarily on
$x^\mu$. It can be proven \cite{OSTRONG} that this parametrization defines ten
independent variables that describe the general spacetime metric. This is a modified version of the well-known parametrization of a metric proposed by Arnowitt \emph{et al.} \cite{ADM} and De Witt \cite{Dewitt} in the context of the canonical formulation of the quantum gravity theory; the off-diagonal metric components $g^{0i} = K^i/V^2$ are related to the effects of rotation. The parametrization (\ref{LT}) is general and covers any Riemannian and Riemann-Cartan spacetimes.

The preferable choice of the tetrad \cite{OST} is the
Schwinger gauge
\begin{equation}\begin{array}{c}\label{coframe}
e_\mu^{\,\widehat{0}} = V\,\delta^{\,0}_\mu,~~~ e_\mu^{\widehat{i}} =
W^{\widehat i}{}_j\left(\delta^j_\mu -
K^j\,\delta^{\,0}_\mu\right),\\
e_{\,\widehat{0}}^\mu = {\frac 1V}\left(\delta_{\,0}^\mu +
\delta_{\,i}^\mu K^i\right), ~~~ e^\mu_{\widehat{i}} =
\delta_{\,j}^\mu W^j{}_{\widehat i},
\end{array}\end{equation}
where the inverse $3\times 3$ matrix, $W^i{}_{\widehat
k}W^{\widehat k}{}_j = \delta_j^i$, is introduced.
The Schwinger gauge is characterized by the conditions
$e_i^{\,\widehat{0}} =0,~ e_{\widehat{i}}^0 = 0$.

For the general metric (\ref{LT}) with the tetrad
(\ref{coframe}) we find explicitly \cite{OSTRONG}
\begin{eqnarray}
\Gamma_{\mu\,\widehat{i}\widehat{0}} &=& {\frac
{1}V}\,W^j{}_{\widehat{i}} \,\partial_jV\,e_\mu{}^{\widehat{0}} -
{\frac 1V}\,{\cal Q}_{(\widehat{i}\widehat{j})}
\,e_\mu{}^{\widehat{j}},\label{connection1}\\
\Gamma_{\mu\,\widehat{i}\widehat{j}} &=& {\frac 1V}\,{\cal
Q}_{[\widehat{i} \widehat{j}]}\,e_\mu{}^{\widehat{0}} + \left({\frak C}_{\widehat{i}\widehat{j} \widehat{k}} + {\frak C}_{\widehat{i}\widehat{k}\widehat{j}} + {\frak C}_{\widehat{k}\widehat{j}\widehat{i}}\right)
e_\mu{}^{\widehat{k}},\label{connection2}
\end{eqnarray}
where 
\begin{eqnarray}
{\cal Q}_{\widehat{i}\widehat{j}} &=&
g_{\widehat{i}\widehat{k}}W^l{}_{\widehat{j}} \left(\dot{W}^{\widehat k}{}_l + K^m\partial_m{W}^{\widehat k}{}_l +
{W}^{\widehat k}{}_m\partial_lK^m\right),\label{Qab}\\
{\frak C}_{\widehat{i}\widehat{j}}{}^{\widehat{k}} &=&
W^l{}_{\widehat{i}}
W^m{}_{\widehat{j}}\,\partial_{[l}W^{\widehat{k}}{}_{m]}=- {\frak C}_{\widehat{j}\widehat{i}}{}^{\widehat{k}},\qquad {\frak C}_{\widehat{i} \widehat{j}\widehat{k}} =
g_{\widehat{k}\widehat{l}}\,{\frak C}_{\widehat{i}
\widehat{j}}{}^{\widehat{l}}.\label{Cabc}
\end{eqnarray}
The dot $\dot{\,}$ denotes the derivative with respect to the time
$t=x^0$. Here ${\frak C}_{\widehat{i}\widehat{j}}{}^{\widehat{k}}$ is nothing but the
anholonomity object for the spatial triad ${W}^{\widehat i}{}_j$.
The indices (which all run from 1 to 3) are raised and lowered with
the help of the spatial part of the flat Minkowski metric
$g_{ab} ={\rm diag}(1, -1, -1,
-1),~g_{\widehat{i}\widehat{j}}= -\delta_{ij}$.

In Riemann-Cartan gravity, the connection (\ref{eqnnn}) should be added by a contribution of a spacetime torsion and takes the form 
\begin{equation}
\Gamma_{\mu\, ab} = \frac12e_\mu^c\left(-C_{cab}+C_{abc}-C_{bca}\right) - K_{\mu\, ab}.\label{GGK}
\end{equation}
The post-Riemannian contortion tensor is given by \cite{OSTORSION}
\begin{eqnarray} K_{\mu\, ab} =-K_{\mu\, ba} = {\frac 12}\left(- T_{\mu ab}+ T_{ab\mu}-T_{b\mu a}\right), \nonumber\\
T_{\mu\nu a} =-T_{\nu\mu a} = e_{a\nu,\mu} -  e_{a\mu,\nu} + \Gamma_{\mu ba} e^b_\nu - \Gamma_{\nu ba} e^b_\mu. \label{Tij}
\end{eqnarray}

To calculate the contribution of the spacetime torsion, it is convenient to use the components of the axial torsion vector
\begin{equation}\label{Taxial}
\check{T}^a = -\,{\frac 12}\,\eta^{abcd}T_{bcd},
\end{equation}
where $\eta^{abcd}$ is the totally antisymmetric Levi-Civita tensor ($\eta_{\hat{0}\hat{1}\hat{2}\hat{3}} = -\eta^{\hat{0}\hat{1}\hat{2}\hat{3}} = +1$).

A direct check shows \cite{Ob} that the Hamiltonian form of the initial Dirac equation (\ref{Dirac0}) is characterized by a non-Hermitian
Hamiltonian. To avoid this difficulty, one can define a new wave function as follows \cite{Ob,OSTRONG,OSTORSION}:
\begin{equation}
\Psi = \left({\sqrt{-g}e_{\widehat{0}}^0}\right)^{1/2}\,\psi.\label{newpsi}
\end{equation}
This form of the nonunitary transformation operator is universal and is applicable to any Riemannian and Riemann-Cartan spacetimes.

For both the Riemannian and Riemann-Cartan spacetimes, the Hermitian Hamiltonian is given by \cite{OSTORSION}
\begin{equation}\begin{array}{c}
{\cal H} = \beta mV + eA_0 + {\frac 1 2}\alpha^{\widehat{i}}\{{\cal F}^j{}_{\widehat{i}},\pi_j\} \\
 +{\frac 12}\left(\bm{K}\cdot\bm{\pi} + \bm{\pi}\cdot\bm{K}\right) +
{\frac 14}\left(\bm{\Xi}\cdot\bm{\Sigma} - \Upsilon\gamma_5\right), \\
\Upsilon = - V\epsilon^{\widehat{i}\widehat{j}\widehat{k}}{\frak C}_{\widehat{i}\widehat{j}\widehat{k}} + V\check{T}^{\widehat{0}},~~~
\Xi^{\widehat{i}} = \epsilon^{\widehat{i}\widehat{j}\widehat{k}}\,{\cal Q}_{\widehat{j}\widehat{k}} - V\check{T}^{\widehat{i}},
\end{array}\label{Hamilton1}
\end{equation} where ${\cal F}^j{}_{\widehat{i}}= VW^j{}_{\widehat i},~\gamma_5=-i\gamma^{\widehat{0}}\gamma^{\widehat{1}}\gamma^{\widehat{2}}
\gamma^{\widehat{3}}$, and $\epsilon^{\widehat{i}\widehat{j}\widehat{k}}$
is the three-dimensional totally antisymmetric Levi-Civita tensor ($e^{\widehat{1}\widehat{2}\widehat{3}}=1$).

The conformal transformation 
of the metric parameters has the form
$$\widetilde{V}=O^{-1}V, ~~~\widetilde{W^{\widehat
i}{}_k}=O^{-1}W^{\widehat
i}{}_k, ~~~\widetilde{W^i{}_{\widehat
k}}=OW^i{}_{\widehat
k}, ~~~ \widetilde{\bm K}=\bm K.$$
As a result,
\begin{eqnarray}
\widetilde{C_{abc}}=OC_{abc}, ~~~ \widetilde{{\cal Q}_{\widehat{i}\widehat{j}}}={\cal Q}_{\widehat{i}\widehat{j}},~~~
\widetilde{{\frak C}_{\widehat{i}\widehat{j}\widehat{k}}}=O{\frak C}_{\widehat{i}\widehat{j}\widehat{k}},
~~~ \widetilde{{\cal F}^j{}_{\widehat{i}}}={\cal F}^j{}_{\widehat{i}}. \label{cnnectn}
\end{eqnarray}
The conformal transformation (\ref{conftrf}) does not change the contortion tensor $K_{\mu\, ab}$. In compliance with this, $\widetilde{\check{T}^a}=O\check{T}^a$.

Thus, $\widetilde{\bm{\Xi}}=\bm{\Xi}$ and $\widetilde{\Upsilon}=\Upsilon$. The above-mentioned relations show that the \emph{Hermitian} Dirac Hamiltonian (\ref{Hamilton1}) is invariant relative to the conformal transformation (\ref{conftrf}) for massless particles \cite{Honnefscalar} and relative to the conformal-like transformation (\ref{conflike}) for massive ones.

We can also note that the anholonomic components of the connection (local Lorentz connection \cite{OSTORSION}) $\Gamma_{\mu\, ab}$ remain unchanged by the conformal transformation. These components together with the tetrad $e_\mu^a$ form the Poincar\'{e} gauge potentials \cite{OSTORSION}. Among these potentials, only the tetrad is changed by the conformal transformation.

It is easy to find the conformal symmetry of the FW Hamiltonian. Equation (\ref{Hamilton1}) can be presented in the form
(\ref{eqH}) (with $\rho_3\rightarrow\beta$), where
\begin{equation} {\cal M}=mV, ~~~ {\cal E}=eA_0 + {\frac 12}\left(\bm{K}\cdot\bm{\pi} + \bm{\pi}\cdot\bm{K}\right) +
{\frac 14}\bm{\Xi}\cdot\bm{\Sigma}, ~~~ {\cal
O}={\frac 1 2}\alpha^{\widehat{i}}\{{\cal F}^j{}_{\widehat{i}},\pi_j\}- \Upsilon\gamma_5.
\label{eqddH} \end{equation}

Equation (\ref{eqddH}) shows that
\begin{equation} \widetilde{\cal M}=O^{-1}{\cal M}, ~~~ \widetilde{\cal E}={\cal E}, ~~~ \widetilde{\cal O}={\cal O}.
\label{eqdconf} \end{equation}

The unitary operator of the FW transformation is given by \cite{PRA}
\begin{equation}
U=\frac{\beta\epsilon+\beta {\cal M}-{\cal
O}}{\sqrt{(\beta\epsilon+\beta {\cal M}-{\cal O})^2}}\,\beta, ~~~
\epsilon=\sqrt{{\cal M}^2+{\cal
O}^2}. \label{eq8f}
\end{equation} This operator is invariant relative to the conformal-like transformation (\ref{conflike}). After the first iteration with the operator (\ref{eq8f}), next iterations eliminate residual odd terms.
The final approximate FW Hamiltonian is equal to \cite{PRA}
\begin{equation}\begin{array}{c}
{\cal H}_{FW}=\beta\epsilon+ {\cal E}+\frac 14\left\{\frac{1}
{2\epsilon^2+\{\epsilon,{\cal M}\}},\left(\beta\left[{\cal O},[{\cal O},{\cal
M}]\right]-[{\cal O},[{\cal O},{\cal
F}]]\right)\right\}, ~~~ {\cal F}={\cal E}-i\hbar\frac{\partial}{\partial
t}. \end{array} \label{eqfingn}
\end{equation} Evidently, this Hamiltonian (whose explicit form is obtained in Ref. \cite{OSTORSION}) is also invariant relative to the conformal-like transformation (\ref{conflike}) in the general case.

The wave function of the equation for the \emph{Hermitian} Dirac Hamiltonian,
\begin{equation} \begin{array}{c}i\frac{\partial \Psi} {\partial t}= {\cal H}\Psi,
\end{array}
\label{eqHmltn} \end{equation} satisfies Eq. (\ref{newpsi}). It is invariant relative to the conformal-like transformation. The FW wave function also possesses this property. The use of Eq. (\ref{newpsi}) allows us to obtain the following property of the wave function of the initial Dirac equation (\ref{Dirac0}) relative to the conformal-like transformation:
\begin{equation}
\widetilde{\psi}=O^{3/2}\psi.
\label{Pentr}\end{equation}
Contrary to the conventional conformal invariance, this property is valid for both massive and massless particles.

All properties stated in this section are valid in the presence of electromagnetic interactions.

We can conclude that the previously ascertained similarity between massless scalar and Dirac particles in Riemannian spacetimes \cite{Honnefscalar} exists for any pointlike particles in both the Riemannian and Riemann-Cartan spacetimes and is not violated by electromagnetic interactions.

\section{Summary}

In the present work, new symmetry properties have been found for fundamental pointlike scalar and Dirac particles (Higgs boson and all leptons) in Riemannian and Riemann-Cartan spacetimes. All results are general and have been obtained in the presence of electromagnetic interactions. The KG equation for a pointlike scalar particle in arbitrary $n$-dimensional Riemannian (or Riemann-Cartan) spacetimes has been brought to the Hermitian form (\ref{eqKGH}). This form is useful to derive the general Hamiltonians in the GFV and FW representations. The GFV Hamiltonians (\ref{eq5i}) and (\ref{eq5e}) are exact. The corresponding 
FW Hamiltonians (\ref{eqf}) and (\ref{eqfem}) are approximate. They are expanded in powers of the
Planck constant and are useful when the de Broglie
wavelength is much smaller than the characteristic distance. Nevertheless, these Hamiltonians are rather general. They cover the nonstationary case and can be applied for a relativistic particle in arbitrarily strong gravitational and inertial fields. In the FW Hamiltonians, terms proportional to
the zeroth and first powers of the Planck constant are determined
exactly while higher-order terms are not specified.

New conformal symmetries of the initial and Hermitian forms of the KG equation were ascertained. When the mass is replaced with any quantity $m'$ satisfying the conformal transformation (\ref{eqconfm}), the \emph{changed} equations become conformally invariant. This property defines the conformal symmetries of the conventional and Hermitian KG equations. The latter equation as well as the obtained Hamiltonians in the GFV and FW representations is invariant relative to
the conformal-like transformation (\ref{conflike}).

Corresponding conformal symmetries are also determined for
both massive and massless Dirac particles. The Dirac and FW Hamiltonians are invariant relative to the conformal-like
transformation (\ref{conflike}). This transformation also defines the conformal symmetry of the initial Dirac equation for a massive particle. When $m'$ defined by Eq. (\ref{eqconfm}) is substituted for $m$, the Dirac wave function has the property (\ref{Pentr}).

It has been proven that all conformal symmetries remain unchanged by an inclusion of electromagnetic interactions. Thus, the results obtained in the present study have
allowed us to state the 
general properties of conformal symmetry for pointlike scalar and Dirac particles (Higgs boson and all leptons) in Riemannian and Riemann-Cartan spacetimes in the presence of electromagnetic interactions.

\section*{Acknowledgments}

The work was supported in part by the Belarusian Republican
Foundation for Fundamental Research (Grant No. $\Phi$14D-007) and
by the Heisenberg-Landau program of the German Ministry for
Science and Technology (Bundesministerium f\"{u}r Bildung und
Forschung).


\end{document}